\documentclass{luxenote}
\pdfoutput=1
\DocAuthors{A.~Person \\ B.~Person \\ C.~Person}
\DocEdBoard{D.~Person \\ E.~Person }
\LUXEDocNo{xxx-yyy-zzzz}
\draftversion{0.1}

\bibliography{refs}

\author{John A. Hallford$^{a, b, *}$ on behalf of the LUXE collaboration} 
\affil{$^a$University College London,\\  Gower St., London, United Kingdom}
\affil{$^b$Deutsches Elektronen Synchrotron,\\ Notkestrasse, Hamburg, Germany}
\affil{$^*$Corresponding author: john.hallford.19@ucl.ac.uk}
\begin{document}

\title{Detector Challenges at the LUXE Experiment}

\date{\today} 

\maketitle

\begin{abstract}

The LUXE experiment, currently in design and planning, aims to perform analyses of strong-field quantum electrodynamics interactions by colliding the high-quality high-energy EU.XFEL electron beam with a powerful laser. With the ability to collide laser pulses with bunches of $1.5 \times 10^9$ electrons / $1 \times 10^8$ photons at 1Hz, this high-statistics environment presents an opportunity to probe rare interactions in a new parameter space of a novel regime. To do this requires a unique suite of detectors to measure three types of particles, at highly varying fluxes dependent on laser interaction parameters. The detectors measure electrons, positrons, or photons, and balance sensitivity with high dynamic range and hardness to radiation damage. Presented in brief in this note are the function, design, and reconstruction methods of each of these detectors. 
\end{abstract}

\vspace{6cm}

Presented at the 30th International Symposium on Lepton Photon Interactions at High Energies, hosted by the University of Manchester, 10-14 January 2022.

\newpage

\section{Laser Und XFEL Experiment}

Quantum electrodynamics relies on perturbative expansions around the vacuum solution. The theory is among the most accurate in the history of physics \cite{fine_structure_constant}, yet as a perturbative expansion, in high energy scales or within extremely high external electromagnetic fields, the theory can fail. The case of such a sufficiently intense electromagnetic field is described by the Schwinger limit, or critical field:

\begin{equation}
 E_{Schw.} \equiv m_e^2 c^3 / q_e \hbar = 1.3 \times 10^{18} \hspace{1mm} V m^{-1} 
\end{equation}

Where each common symbol has its usual physical constant meaning, including $q_e$ and $m_e$ as the charge and mass of the electron respectively. This is several orders of magnitude higher than achievable currently in modern laboratories. Yet by using the boosted frame of highly energetic electrons, due to relativistic contraction, the relative field intensity experienced is magnified.

LUXE (Laser Und XFEL Experiment) intends to make use of the most intense electric fields currently feasible in the lab today: highly focussed laser beam pulses, which benefit from the chirped amplification technique \cite{donna_strickland}. By colliding these pulses with a high-energy particle beam, high values of the assisted electric field and therefore $\chi$, a quantum non-linearity parameter, can be observed. A value of $\chi=1$ implies an assisted electric field at the Schwinger limit; LUXE intends to reach $\chi=3-5$. Further discussions of the strong-field QED physics may be found in \cite{Fedotov}. This general idea was performed first at SLAC in the 1990s, making use of the $\sim$~46 GeV SLAC electron beam \cite{slac_e144}. LUXE takes advantage of the developments in laser technology since then, and will use the EU.XFEL beam, of electrons up to 17.5 GeV in energy and with bunch population of $\sim10^9$ - but a much more powerful laser. In LUXE's phase-0, a peak 40 TW laser pulse will be used, before an upgrade during the experiment's lifetime to a $\sim350$ TW laser. Given a 1Hz repetition rate, with a running schedule of several years anticipated, LUXE has the opportunity to gather enough statistics to make high-quality measurements of the strong-field interactions for a range in $\chi$.

This is performed in two modes of operation: in the electron-laser collision mode, the electron bunches directly from the XFEL.EU are intercepted by the laser pulses. In the $\gamma$-laser mode, the electron bunches are used to create a high-energy beam of photons. This is done primarily via a thin heavy-metal target (35 \textmu m of Tungsten), but also there exists the option of splitting the laser to create an Inverse-Compton-Scattered (ICS), effectively monochromatic, photon beam.

In both modes two first-order strong-field interactions are expected, creating resultant particles of three types: electrons, positrons and photons. Figure \ref{fig:feynman} shows the two main interactions. 

\begin{figure}[h!] 
\begin{center}

\begin{subfigure}{0.24\hsize} 
\includegraphics[width=\textwidth]{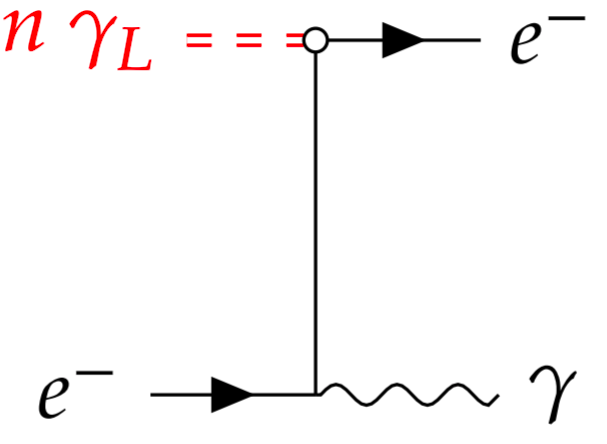}
\caption{  \label{fig:feynman}}
\label{fig:setups}
\end{subfigure}
\hspace{3cm}
\begin{subfigure}{0.24\hsize}
\includegraphics[width=\textwidth]{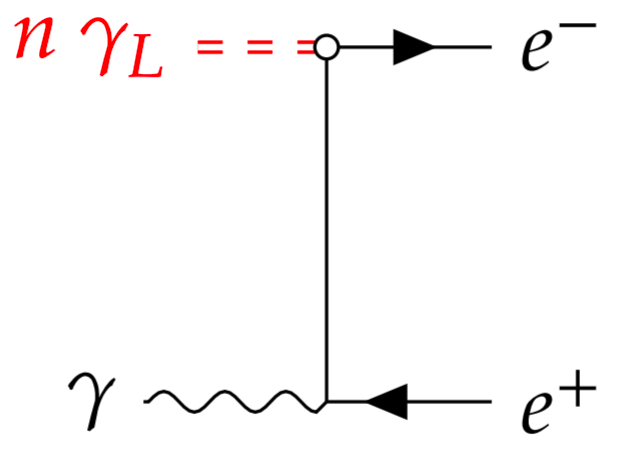}
\caption{ \label{fig:feynman} }
\end{subfigure}
\end{center}

\caption{\label{fig:feynman} Feynman diagrams of the two principal strong-field interactions expected in LUXE. \textbf{(a)} Shows non-linear Compton scattering and \textbf{(b)} shows the multiphoton Breit-Wheeler process. The $\gamma_L$ index denotes any number of laser photons. Both these processes can happen one after another in the interaction point, forming an effective `trident' process.}

\end{figure}

This provides unique challenges in detecting the resulting particles, the only way of reconstructing the strong-field interactions. Figure \ref{fig:setups} shows the array of detectors used in each mode (e-laser or $\gamma$-laser). In particular it shows that magnetic fields are used to separate the outgoing particles by charge, and so each particle type has its own detector system, composed of at least two complementary sub-detectors. 

The previously described interactions exhibit vastly different cross-sections, and so vastly different rates of photons to positrons to electrons. Within the e-laser mode, depending on the laser spot parameters, electrons which have scattered (through non-linear Compton scattering) below the beam-energy (and remain of order 1-15 GeV) are produced at rates of $10^7$ - $10^9$ per beam-laser crossing. Photons from the non-linear Compton scattering, also of order 1-15 GeV, are too produced at around $10^7$ - $10^9$ per beam-laser crossing. Downstream from the initial magnetic field (at magnitude of $\sim1$ T) both these electrons and photons continue with intense flux, and so pose difficulty in reliable detection as they can damage detectors with their high local dose of radiation. Positrons are produced in the e-laser mode at around 5 GeV, at lower rates of $10^{-2}$ - $10^5$ per beam-laser crossing. This must be measured in the context of the wider experiment, where bunches up to $1.5 \times 10^9$ electrons are dumped within the same experimental chamber. This requires significant power to exclude background from signal, when for a low value of $\xi$, pair production is expected only once in 100 collisions. For each positron, an electron is also produced, but this is not typically detectable above the non-linear Compton scattered electron flux. The interaction can be fully reconstructed using only the positron rates/energy distributions due to the symmetry between the electrons \& positrons. The expected rates of these particles have been estimated using the strong-field QED simulation Ptarmigan \cite{ptarmigan}.  

In the $\gamma$-laser mode, the photon beam is of order $10^8$ in bunch population and again around 1-15 GeV when created by bremsstrahlung. If the Inverse Compton Scattering technique is used, the bunch population is orders of magnitude lesser and has energy $\sim9$ GeV. The electrons and positrons are created with equal number and energy distributions, with around 5 GeV in energy in each and varying rates of around $10^{-2}$ - $10^2$ per bunch crossing. The bremsstrahlung emission has been modelled and simulated by Geant4 \cite{geant4}.

\section{Detectors}
Non-linear Compton-scattered electrons are produced in high numbers and are counted and reconstructed with respect to energy, rather than analysed individually, using magnetic fields as spectrometers. A screen of scintillating material is used in this region in conjunction with a segmented Cherenkov detector. The high electron flux induces high scintillation light levels, allowing remote optical CMOS cameras to detect signal at high position resolution, around $\sim$ 100 \textmu m. This then gives a high energy resolution, far below the goal of $\sim 2 \%$ for LUXE electron detection. The scintillator is sensitive to low-energy background radiation, however. In the Cherenkov detector the high flux means a Cherenkov medium with low refractive index (e.g. air) can be used, which provides very few signal photons per-electron, but excludes low energy (background) charged particles E $\textless$ 20 MeV. The relatively low signal per-electron is not a problem as again the incident flux is so large. Reflective straw tubes carry the light produced from the electrons to an array of photodetectors. In comparison to the CMOS camera sensor, these photodetectors also allow a greater dynamic range.   

The scintillator \& Cherenkov setup is also replicated to detect electrons which radiate photons to be used in the $\gamma$-laser mode. As this electron spectrum cannot be fully covered, especially the crucial low-energy region which corresponds to high-energy photons, this site is used mainly for shot-to-shot monitoring. The photon detection system downstream is the primary measurement of the photon beam which has been generated.  


\begin{figure}[h!] 
\begin{center}
\begin{subfigure}{0.49\hsize} 
\includegraphics[width=\textwidth]{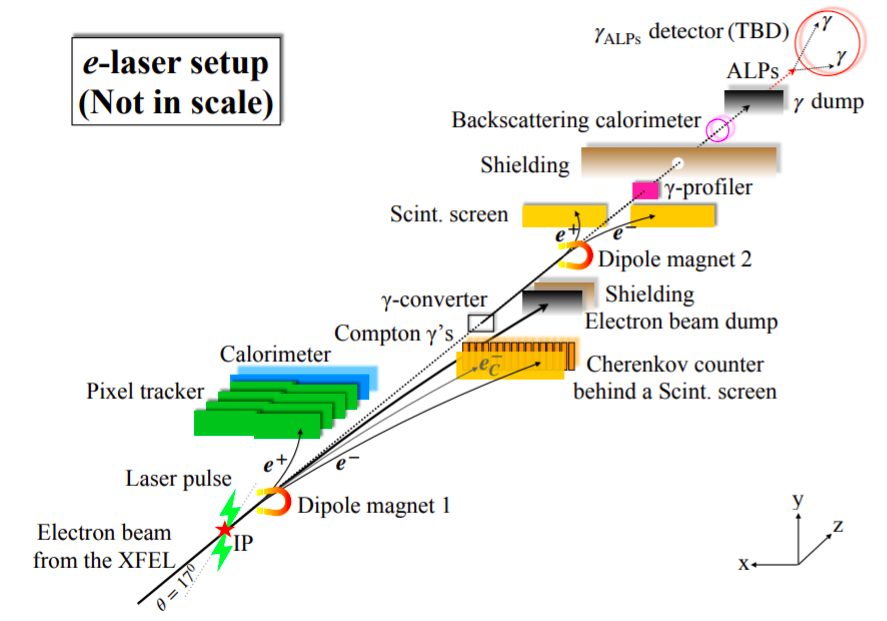}
\caption{  \label{fig:setups}}
\label{fig:setups}
\end{subfigure}
\begin{subfigure}{0.49\hsize}
\includegraphics[width=\textwidth]{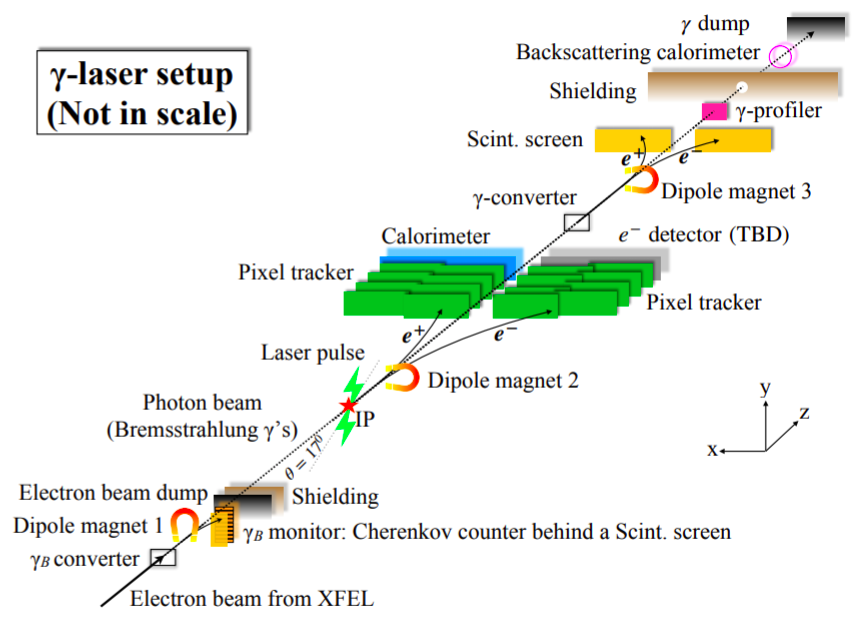}
\caption{ \label{fig:setups} }
\end{subfigure}
\end{center}
\caption{\label{fig:setups} A schematic of the LUXE experiment in its two modes. \textbf{(a)} is the e-laser mode and \textbf{(b)} the $\gamma$-laser mode.}

\end{figure}

The low rates expected for the Breit-Wheeler process means there must be a high power of background exclusion for positron detection. For this the use of technologies which can utilise track reconstruction - including direct analysis of track calorimetry - are required. The combination of silicon pixel-trackers and silicon/GaAs sampling calorimeter(s) has been chosen for this. In front sits the silicon tracker, composed of four layers of two staves each of 25 \textmu m of silicon. The design is similar to (and based on) the ALPIDE tracker in development for ALICE \cite{alpide}. A Kalman filter fitting algorithm is used to identify tracks, and those which fit geometric expectations for the charged particles can be identified as signal, and so background excluded \cite{kalman}. To assist with general background, mitigation shielding is placed above and below the beampipe which separates the high-flux electron side from the positron side. Quantum computing solutions are also being investigated to try improve on background rejection and so achieve further significance even with very low positron rates. The sampling calorimeter uses 20 tungsten plates to induce showering within the detector, then sampled by silicon or GaAs sensor layers to reconstruct energy and position of tracks. This destructively but directly measures the energy of the track. Performance for this detector, from simulations of the LumiCal prototype \cite{lumical}, offer energy resolution of $\sigma / E = 20 \% /\sqrt{E(GeV)}$ and position resolution at 500 \textmu m.

\begin{figure}[t!] 
\begin{center}
\begin{subfigure}{0.49\hsize} 
\includegraphics[width=\textwidth]{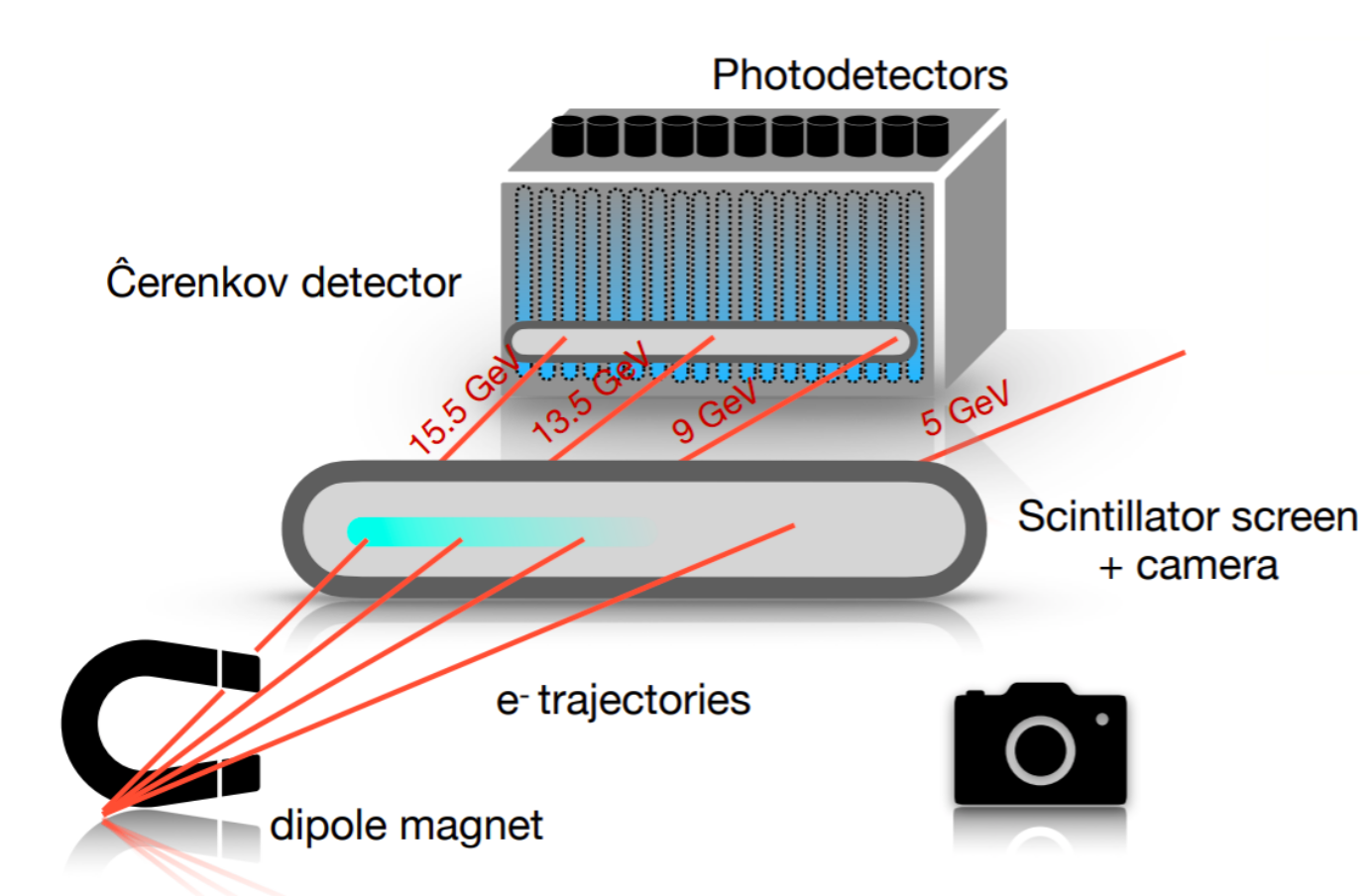}
\caption{  \label{fig:detectors}}
\label{fig:detectors}
\end{subfigure}
\begin{subfigure}{0.49\hsize}
\includegraphics[width=\textwidth]{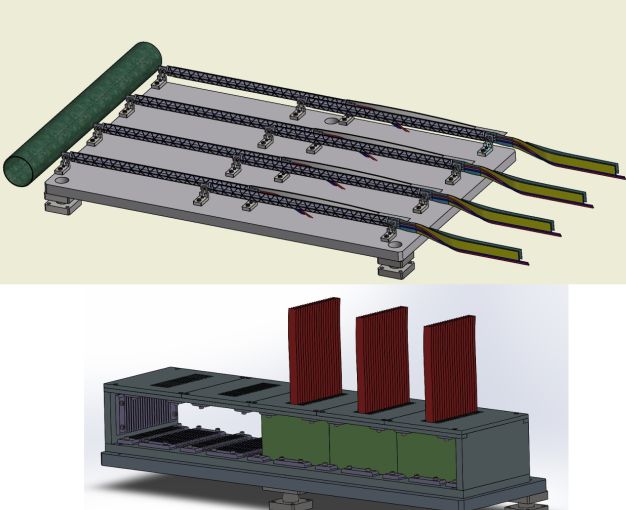}
\caption{ \label{fig:detectors} }
\end{subfigure}
\begin{subfigure}{0.49\hsize} 
\includegraphics[width=\textwidth]{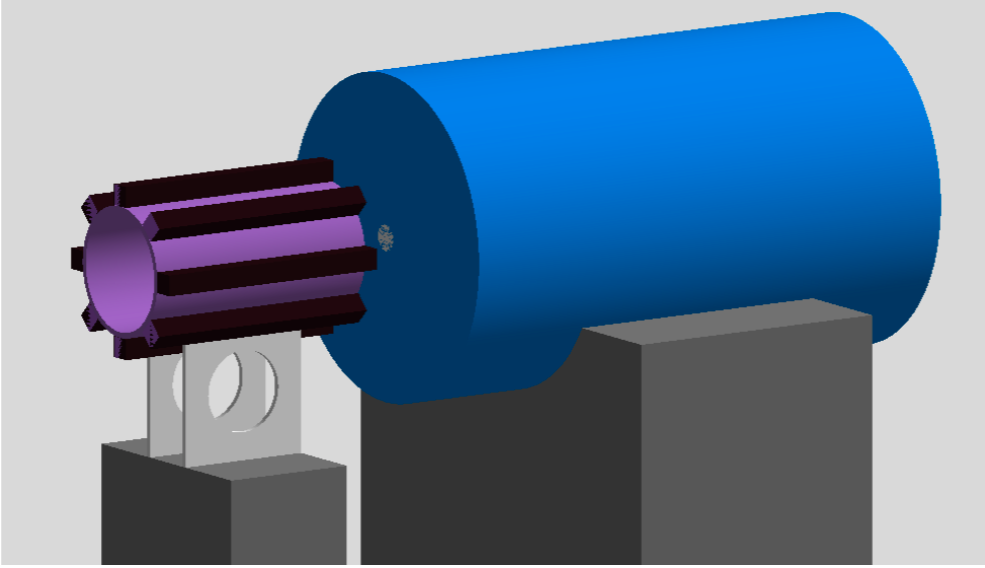}
\caption{  \label{fig:detectors}}
\label{fig:detectors}
\end{subfigure}
\begin{subfigure}{0.49\hsize}
\includegraphics[width=\textwidth]{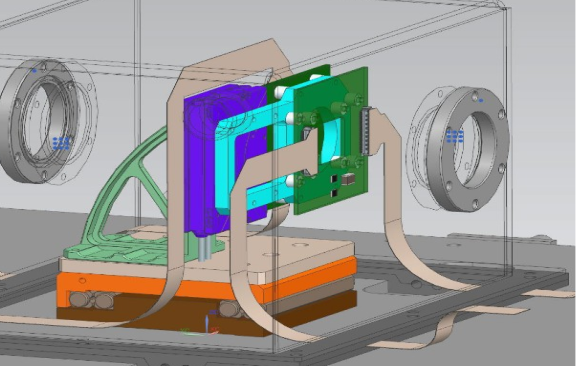}
\caption{ \label{fig:detectors} }
\end{subfigure}
\end{center}
\caption{\label{fig:detectors} Schematics of various LUXE detectors. \textbf{(a)} Shows the electron detection system of both the scintillation screen \& camera system and Cherenkov detector \textbf{(b)} (above) An illustration of the silicon-wafer tracker detector and the electromagnetic calorimeter (below). \textbf{(c)} shows the lead-glass gamma flux calorimeter blocks in purple, around the beampipe and in front of the photon beam-dump. \textbf{(d)} Shows an illustration of the sapphire-based gamma profiler.}

\end{figure}

Three independent detectors are used to measure various characteristics of the gamma beam in the photon detection system. The gamma spectrometer uses a thin tungsten target to convert a proportion ($\sim$1\%) of the gamma beam to electron-positron pairs, and using a Bethe-Heitler deconvolution algorithm and the summed energy of the $e^-/e^+$ pair, the gamma beam is reconstructed with energy distribution \cite{Fleck:2020}. The energy profile of the electrons and positrons are reconstructed with another scintillation screen \& camera system, similar to the electron detection system, although the camera models used are more sensitive, with lower typical electronic noise, as the signal flux is far lower.  A sapphire-microstrip gamma beam profiler measures the flux and physical shape of the produced gamma beam. Two layers of these thin strips, one arrayed in the x dimension and one in the y, respond to the charge flux in each of these directions to a resolution of 5 \textmu m. Sapphire is chosen for its resistance to radiation damage. This shape is crucial as it informs us immediately about the quality of the particle-laser interaction from bunch-to-bunch.  Finally, in front of the gamma beam dump, a backscattering calorimeter measures the total photon energy flux using photon back-scatters from the photon-beam dump. Eight lead-glass blocks are arranged around the beampipe; electromagnetic showering within the glass leads to Cherenkov light which is detected by photomultiplier tubes.

Schematics and illustrations of selected detectors can be seen in figure \ref{fig:detectors}. Further reading, including the prospect of Beyond the Standard Model beam-dump searches for Axion-like particles, is present within the published LUXE Conceptual Design Report \cite{CDR}.


\printbibliography

@article{CDR,
   title={Conceptual design report for the LUXE experiment},
   volume={230},
   ISSN={1951-6401},
   url={http://dx.doi.org/10.1140/epjs/s11734-021-00249-z},
   DOI={10.1140/epjs/s11734-021-00249-z},
   number={11},
   journal={The European Physical Journal Special Topics},
   publisher={Springer Science and Business Media LLC},
   author={Abramowicz, H. and Acosta, U. and Altarelli, M. and Aßmann, R. and Bai, Z. and Behnke, T. and Benhammou, Y. and Blackburn, T. and Boogert, S. and Borysov, O. and et al.},
   year={2021},
   month={Sep},
   pages={2445–2560}
}

@article{lumical,
      author         = "Abramowicz, H. and others",
      title          = "{Performance and Moli\`ere radius measurements using a
                        compact prototype of LumiCal in an electron test beam}",
      year           = "2019",
      eprint         = "1812.11426",
      archivePrefix  = "arXiv",
      primaryClass   = "physics.ins-det",
      journal        = "Eur. Phys. J.",
      volume         = " C 79",
      pages          = "579",
      SLACcitation   = "%%CITATION = ARXIV:1812.11426;%%"
}

@article{kalman,
		title = "Simultaneous pattern recognition and track fitting by the Kalman filtering method",
		journal = "Nuclear Instruments and Methods in Physics Research Section A: Accelerators, Spectrometers, Detectors and Associated Equipment",
		volume = "294",
		number = "1",
		pages = "219",
		year = "1990",
		issn = "0168-9002",
		doi = "https://doi.org/10.1016/0168-9002(90)91835-Y",
		url = "http://www.sciencedirect.com/science/article/pii/016890029091835Y",
		author = "P. Billoir and S. Qian"
}

@article{alpide,
      author         = "Mager, M.",
      title          = "{ALPIDE, the Monolithic Active Pixel Sensor for the ALICE
                        ITS upgrade}",
      booktitle      = "{Proceedings, 13th Pisa Meeting on Advanced Detectors :
                        Frontier Detectors for Frontier Physics (FDFP 2015): La
                        Biodola, Isola d'Elba, Italy, May 24-30, 2015}",
      collaboration  = "ALICE",
      journal        = "Nucl. Instrum. Meth.",
      volume         = "A824",
      year           = "2016",
      pages          = "434",
}

@article{donna_strickland,
title = "{Compression of Amplified Chirped Optical Pulses}",
journal = "Optics Communications",
volume = "56",
number = "3",
pages = "219 - 221",
year = "1985",
issn = "0030-4018",
doi = "https://doi.org/10.1016/0030-4018(85)90120-8",
url = "http://www.sciencedirect.com/science/article/pii/0030401885901208",
author = "D. Strickland and G. Mourou"
}

@article{fine_structure_constant,
  title = "{New Determination of the Fine Structure Constant and Test of the Quantum Electrodynamics}",
  author = {Bouchendira, R. and Clad\'e, Pierre and Guellati-Kh\'elifa, Sa\"{\i}da and Nez, Fran\ifmmode \mbox{\c{c}}\else \c{c}\fi{}ois and Biraben, Fran\ifmmode \mbox{\c{c}}\else \c{c}\fi{}ois},
  journal = {Phys. Rev. Lett.},
  volume = {106},
  issue = {8},
  pages = {080801},
  numpages = {4},
  year = {2011},
  month = {Feb},
  publisher = {American Physical Society},
  doi = {10.1103/PhysRevLett.106.080801},
  url = {https://link.aps.org/doi/10.1103/PhysRevLett.106.080801}
}

@article{slac_e144,
  title = "{Studies of nonlinear QED in collisions of 46.6 GeV electrons with intense laser pulses}",
  author = {Bamber, C. and Boege, S. J. and Koffas, T. and Kotseroglou, T. and Melissinos, A. C. and Meyerhofer, D. D. and Reis, D. A. and Ragg, W. and Bula, C. and McDonald, K. T. and Prebys, E. J. and Burke, D. L. and Field, R. C. and Horton-Smith, G. and Spencer, J. E. and Walz, D. and Berridge, S. C. and Bugg, W. M. and Shmakov, K. and Weidemann, A. W.},
  journal = {Phys. Rev. D},
  volume = {60},
  issue = {9},
  pages = {092004},
  numpages = {43},
  year = {1999},
  month = {Oct},
  publisher = {American Physical Society},
  doi = {10.1103/PhysRevD.60.092004},
  url = {https://link.aps.org/doi/10.1103/PhysRevD.60.092004}
}

@article{Fedotov,

    author = "Fedotov, A. and Ilderton, A. and Karbstein, F. and King, B. and Seipt, D. and Taya, H. and Torgrimsson, G.",

    title = "{Advances in QED with intense background fields}",

    eprint = "2203.00019",

    archivePrefix = "arXiv",

    primaryClass = "hep-ph",

    reportNumber = "RIKEN-iTHEMS-Report-22",

    month = "2",

    year = "2022"

}

@article{geant4,
title = "{Geant4: a Simulation Toolkit}",
journal = "Nuclear Instruments and Methods in Physics Research Section A: Accelerators, Spectrometers, Detectors and Associated Equipment",
volume = "506",
number = "3",
pages = "250 - 303",
year = "2003",
issn = "0168-9002",
doi = "https://doi.org/10.1016/S0168-9002(03)01368-8",
url = "http://www.sciencedirect.com/science/article/pii/S0168900203013688",
author = "S. Agostinelli and J. Allison and K. Amako and J. Apostolakis and H. Araujo and P. Arce and M. Asai and D. Axen and S. Banerjee and G. Barrand and F. Behner and L. Bellagamba and J. Boudreau and L. Broglia and A. Brunengo and H. Burkhardt and S. Chauvie and J. Chuma and R. Chytracek and G. Cooperman and G. Cosmo and P. Degtyarenko and A. Dell'Acqua and G. Depaola and D. Dietrich and R. Enami and A. Feliciello and C. Ferguson and H. Fesefeldt and G. Folger and F. Foppiano and A. Forti and S. Garelli and S. Giani and R. Giannitrapani and D. Gibin and J.J. {Gómez Cadenas} and I. González and G. {Gracia Abril} and G. Greeniaus and W. Greiner and V. Grichine and A. Grossheim and S. Guatelli and P. Gumplinger and R. Hamatsu and K. Hashimoto and H. Hasui and A. Heikkinen and A. Howard and V. Ivanchenko and A. Johnson and F.W. Jones and J. Kallenbach and N. Kanaya and M. Kawabata and Y. Kawabata and M. Kawaguti and S. Kelner and P. Kent and A. Kimura and T. Kodama and R. Kokoulin and M. Kossov and H. Kurashige and E. Lamanna and T. Lampén and V. Lara and V. Lefebure and F. Lei and M. Liendl and W. Lockman and F. Longo and S. Magni and M. Maire and E. Medernach and K. Minamimoto and P. {Mora de Freitas} and Y. Morita and K. Murakami and M. Nagamatu and R. Nartallo and P. Nieminen and T. Nishimura and K. Ohtsubo and M. Okamura and S. O'Neale and Y. Oohata and K. Paech and J. Perl and A. Pfeiffer and M.G. Pia and F. Ranjard and A. Rybin and S. Sadilov and E. {Di Salvo} and G. Santin and T. Sasaki and N. Savvas and Y. Sawada and S. Scherer and S. Sei and V. Sirotenko and D. Smith and N. Starkov and H. Stoecker and J. Sulkimo and M. Takahata and S. Tanaka and E. Tcherniaev and E. {Safai Tehrani} and M. Tropeano and P. Truscott and H. Uno and L. Urban and P. Urban and M. Verderi and A. Walkden and W. Wander and H. Weber and J.P. Wellisch and T. Wenaus and D.C. Williams and D. Wright and T. Yamada and H. Yoshida and D. Zschiesche"
}

@misc{ptarmigan,
 author="T. G. Blackburn",
 howpublished = "\url{https://github.com/tgblackburn/ptarmigan}"
}

@article{Fleck:2020,
    author = "K. Fleck and N. Cavanagh and G. Sarri",
    title = "{Conceptual Design of a High-flux Multi-GeV Gamma-ray Spectrometer}",

    doi = "https://doi.org/10.1038/s41598-020-66832-x",
    journal = "Sci. Rep.",
    volume = "10",
    number = "",
    pages = "9894",
    year = "2020"
}

\end{document}